\documentclass[pra,showpacs,showkeys,amsfonts,amsmath]{revtex4}
\usepackage{bm}
\usepackage{graphicx}
\RequirePackage{mathptm}

\numberwithin{equation}{section}
\newcommand{\si}[1]{\sigma_{#1}}
\newcommand{\sa}[2]{\sigma_{#1}^{#2}}
\newcommand{\ip}[2]{\langle \,{#1},\,{#2}\,\rangle}
\newcommand{\Wasy}[4]{\begin{cases}
#1 ,&#2\\
#3 ,&#4
\end{cases}}
\newcommand{\ro}{\rho}

\newcommand{\Ga}{\Gamma}
\newcommand{\om}{\omega_{0}}

\newcommand{\I}{\mathbb I}
\newcommand{\ket}[1]{|{#1}\rangle}
\newcommand{\bra}[1]{\langle {#1} |}

\newcommand{\C}{\mathbb C}

\newcommand{\tr}{\mathrm{tr}\,}

\newcommand{\mr}[1]{\mathrm{#1}}
\newcommand{\kw}{a^{2}}
\newcommand{\aaa}{1-a^{2}}
\newcommand{\DS}{\displaystyle}
\begin{document}
\title{Generation of Werner states and preservation of entanglement in a noisy environment}
\author{Lech Jak{\'o}bczyk\footnote{Corresponding author.
E-mail addres: ljak@ift.uni.wroc.pl}}
\affiliation{Institute of Theoretical Physics\\ University of
Wroc{\l}aw\\
Pl. M. Borna 9, 50-204 Wroc{\l}aw, Poland}
\author{Anna Jamr{\'o}z}
\affiliation{Institute of Theoretical Physics\\ University of
Wroc{\l}aw\\
Pl. M. Borna 9, 50-204 Wroc{\l}aw, Poland}
\begin{abstract}
We study the influence of noisy environment on the evolution of
two-atomic system in the presence of collective damping.
Generation of Werner states as asymptotic stationary states of
evolution is described. We also show that for some initial states
the amount of entanglement is preserved during the evolution.
\end{abstract}
\maketitle
\section{Introduction}
Entanglement of quantum states is the most non-classical feature
of quantum systems and one of the key resources in quantum
information theory \cite{NC}. In real quantum systems, inevitable
interactions with surrounding environment may lead to decoherence
resulting in degradation of entanglement. Since entanglement once
it has been lost, cannot be restored by local operations, it is
important to understand the process of disentanglement to control
the effects of noise and to preserve as much entanglement as
possible. The main motivation of the  investigations presented in
the paper is to study the possibility of the preservation of
entanglement in the model of two two-level atoms interacting with
environment with maximal noise (the case of thermal noise will be
discussed elsewhere). In the Markovian approximation, the
influence of that kind of environment on a single atom is
described by dynamical semi-group with Lindblad generator in which
the transition rates from ground state to excited state of the
atom and vice versa, are equal to decay rate $\Ga_{0}$. This noisy
dynamics is related to the following limiting procedure:
temperature of thermal reservoir tends to infinity, whereas the
coupling strength goes to zero \cite{CMB,Y}. On the other hand,
the collective properties of two - atomic systems can alter the
decay process compared with the single atom. It was already shown
by Dicke \cite{Dicke} in the case of spontaneous emission and
environment in the vacuum state that there are states with
enhanced emission rates (superradiant states) and such that the
emission rate is reduced (subradiant states). In the latter case,
two-atom system can decohere slower compared with individual atoms
and some amount of initial entanglement can be preserved or even
created by the indirect interaction between atoms \cite{J, FT1,
FT2,JJ}.
\par
In the model discussed in the paper, the collective dynamics of
two atoms is described by damping rate $\Ga$ which depends on
interatomic distance. When the atoms are separated by a large
distance, one can assume that they are located inside two
independent environments and $\Ga =0$. In that case, the noisy
dynamics brings all initial states into unique asymptotic state
which is maximally mixed. Moreover, all entangled states
disentangle in finite time \cite{JJ1}. When atoms are confined in
a region smaller than the radiation wave length, the collective
damping rate is close to the decay rate, so we can use the
approximation $\Ga=\Ga_{0}$. In that case, similarly as in the
Dicke model, the antisymmetric state (singlet state) is decoupled
from the environment and therefore is stable. This is the main
physical reason why entanglement can be preserved in spite of the
the influence of noisy environment.
\par
In this paper we are mainly interested in robust entanglement, so
we study long time (asymptotic) behaviour of dynamical semi -
group. We also not discuss here the case of arbitrary separation
of atoms, since in that case  all states disentangle
asymptotically. Small distance separation modeled by the condition
$\Ga=\Ga_{0}$, leads to the interesting semi - group which is not
ergodic: asymptotic stationary states depend on initial conditions
and can be parametrized by fidelity of initial state with respect
to singlet state. We show that some of these asymptotic states are
entangled. They belong to the important class of Werner states
\cite{W}. In particular, we prove that if fidelity is greater then
$1/2$, the asymptotic Werner state is entangled. So collective
damping can produce correlations between atoms which partially
overcome the effect of decoherence, but in contrast to the zero
temperature case \cite{J,FT1,FT2,JJ}, this process cannot create
entanglement. Our noisy dynamics has also another remarkable
property: there are initial entangled states for which
entanglement is preserved during the evolution, although the
process of decoherence takes place, resulting in decreasing of
purity. As we prove, asymptotic entanglement depends only on the
overlap of the initial state with singlet state but not on its
entanglement. So initial states with the same entanglement can
behave differently with respect to the noise. This opens the
possibility of protecting some entanglement of the initial state
by performing local operations (which do not change entanglement)
to maximize its overlap with singlet state.
\par
When the interatomic separation is small, coupling by the dipol --
dipol interaction plays a significant role. It causes the
entanglement between two atoms to oscillate in time, so for some
period initial entanglement may even increase, but the noise
decreases the amplitude of these oscillations, and asymptotically
its contribution to the preservation of entanglement vanishes.
Since we mainly study here the role of noise in the time evolution
of entanglement, we are not discussing in detail the interesting
problem of the influence of dipol -- dipol coupling on this
evolution. Similarly, the experimental side of the problems
studied here is beyond the scope of the present paper, where we
investigate the theoretical model of the compound system
interacting with maximally noised environment.
\par
\section{Noisy dynamics of two-atomic system }
Time evolution of a density matrix of two two-level atoms $A$ and
$B$ interacting with a noisy environment of the type discussed in
the introduction, can be described by the following master equation
\begin{equation}
\frac{d\ro}{dt}=-i[H,\ro]\,+\,L_{\mr{N}}\ro \label{mastereq}
\end{equation}
Here
\begin{equation}
H=\om \sum\limits_{j=A,B} \sa{3}{j}\;+\;\sum\limits_{j,k=A,B,\\
j\neq k}\Omega_{jk}\sa{+}{j}\sa{-}{k} \label{hamiltonian}
\end{equation}
and
\begin{eqnarray}
L_{\mr{N}}\ro=\frac{1}{2}\,\sum\limits_{j,k=A,B}\;\Ga_{jk}\,
(2\sa{+}{j}\ro\sa{-}{k}-\sa{-}{j}\sa{+}{k}\ro-\ro\sa{-}{j}\sa{+}{k}
+2\sa{-}{j}\ro\sa{+}{k}-\sa{+}{j}\sa{-}{k}\ro-\ro\sa{+}{j}\sa{-}{k})\label{gener}
\end{eqnarray}
where
$$
\sa{\pm}{A}=\si{\pm}\otimes\I,\quad
\sa{\pm}{B}=\I\otimes\si{\pm},\quad \sa{3}{A}=\si{3}\otimes \I,\quad
\sa{3}{B}=\I\otimes \si{3},\quad \si{\pm}=\frac{\DS 1}{\DS
2}\,(\si{1} \pm \,i \si{2})
$$
and we identify ground state $\ket{0}$ and excited state $\ket{1}$
of the atom $A$ or $B$ with vectors $\left(\begin{array}{c}
 0\\
 1\\
\end{array}\right)$ and $ \left(\begin{array}{c}
1\\
0\\
\end{array}\right)$ in $\C^{2}$. In the hamiltonian term
(\ref{hamiltonian}), $\om$ is the frequency of the transition
$\ket{0}\to \ket{1}$ and $\Omega_{AB}=\Omega_{BA}=\Omega$
describes interatomic coupling by the dipol-dipol interaction. On
the other hand, noisy dynamics is given by the generator
(\ref{gener}) with
\begin{equation}
\Ga_{AA}=\Ga_{BB}=\Ga_{0}\label{rateAA}
\end{equation}
and
\begin{equation}
\Ga_{AB}=\Ga_{BA}=\Ga \label{rateAB}
\end{equation}
In the equality (\ref{rateAA}), $\Ga_{0}$ is a decay rate of
individual atom. The parameter $\Ga$ in the equality
(\ref{rateAB}) describes the collective damping rate of two atoms
interacting with a noisy environment. In our model, $\Ga$
satisfies
$$
\Ga=g(R)\,\Ga_{0}
$$
where $g(R)$ is the function of the distance $R$ between atoms
such that $g(R)\to 1$ when $R\to 0$. Notice that (\ref{gener}) can
be rewritten as
\begin{eqnarray}
L_{\mr{N}}\ro=&& \Ga_{0}\,\left[\,
\sa{+}{A}\ro\sa{-}{A}+\sa{+}{B}\ro\sa{-}{B}+\sa{-}{A}\ro\sa{+}{A}
+\sa{-}{B}\ro\sa{+}{B}-2\ro\,\right]\nonumber \\
&&+ \frac{1}{2}\, \Ga\, \left[\,
2\sa{+}{A}\ro\sa{-}{B}+2\sa{-}{A}\ro\sa{+}{B}-\sa{-}{A}\sa{+}{B}\ro-\ro\sa{-}{A}\sa{+}{B}
-\sa{+}{A}\sa{-}{B}\ro-\ro\sa{+}{A}\sa{-}{B}\,\right]\nonumber \\
&&+ \frac{1}{2}\, \Ga\, \left[\,
2\sa{+}{B}\ro\sa{-}{A}+2\sa{-}{B}\ro\sa{+}{A}-\sa{-}{B}\sa{+}{A}\ro-\ro\sa{-}{B}\sa{+}{A}
-\sa{+}{B}\sa{-}{A}\ro-\ro\sa{+}{B}\sa{-}{A}\,\right]\label{genszum}
\end{eqnarray}
\par
The master equation (\ref{mastereq}) describing the time evolution
of a two-atomic system in a noisy environment can be used to
obtain the equations for matrix elements of any density matrix. To
simplify calculations one can work in the basis of so called
collective states in the Hilbert space $\C^{4}$ \cite{FT}. If
\begin{equation}
f_{1}=\ket{1}\otimes\ket{1},\, f_{2}=\ket{1}\otimes\ket{0},\,
f_{3}=\ket{0}\otimes\ket{1},\,
f_{4}=\ket{0}\otimes\ket{0}\label{canbasis}
\end{equation}
then this basis containing excited state, ground state  and
symmetric and antisymmetric combination of the product states, is
defined as follows
\begin{equation}
\ket{e}=f_{1},\, \ket{g}=f_{4},\, \ket{s}=\frac{\DS 1}{\DS
\sqrt{2}}\,(f_{2}+f_{3}),\, \ket{a}=\frac{\DS 1}{\DS
\sqrt{2}}\,(f_{2}-f_{3})
\end{equation}
In the basis of collective states, two--atom system can be treated
as single four -- level system with ground state $\ket{g}$, excited
state $\ket{e}$ and two intermediate states $\ket{s}$ and $\ket{a}$.
From (\ref{mastereq}) it follows that the matrix elements with
respect to the basis $\ket{e},\, \ket{s},\, \ket{a},\, \ket{g}$ of
the state $\ro$ satisfy
\begin{subequations}
\label{eq}
\begin{eqnarray}
\frac{d\ro_{aa}}{dt}&&=-(\Ga_{0}-\Ga)\,(2\ro_{aa}-\ro_{gg}-\ro_{ee})\label{eq1}\\
\frac{d\ro_{ss}}{dt}&&=-(\Ga_{0}+\Ga)\,
(2\ro_{ss}-\ro_{gg}-\ro_{ee})\label{eq2}\\
\frac{d\ro_{ee}}{dt}&&=-2\Ga_{0}\ro_{ee}+(\Ga_{0}+\Ga)\ro_{ss}+
(\Ga_{0}-\Ga)\ro_{aa}\label{eq3}\\
\frac{d\ro_{gg}}{dt}&&=-2\Ga_{0}\ro_{gg}+
(\Ga_{0}+\Ga)\ro_{ss}+(\Ga_{0}-\Ga)\ro_{aa}\label{eq4}\\
\frac{d\ro_{eg}}{dt}&&=-(2\Ga_{0}+4i\om)\ro_{eg}\label{eq5}\\
\frac{d\ro_{as}}{dt}&&=-(2\Ga_{0}-2i\Omega)\ro_{as}\label{eq6}\\
\frac{d\ro_{ae}}{dt}&&=-(\Ga_{0}-\Ga)\ro_{gs}+\left[\,
(\Ga+i\Omega)-2(\Ga_{0}-i\om)\,\right]\ro_{ae}\label{eq7}\\
\frac{d\ro_{ag}}{dt}&&=-(\Ga_{0}-\Ga)\ro_{ea}+\left[\,
(\Ga+i\Omega)-2(\Ga_{0}+i\om)\,\right]\ro_{ag}\label{eq8}\\
\frac{d\ro_{se}}{dt}&&=(\Ga_{0}+\Ga)\ro_{gs}-\left[\, (\Ga
+i\Omega)+2(\Ga_{0}-2i\om)\,\right]\ro_{se}\label{eq9}\\
\frac{d\ro_{sg}}{dt}&&=(\Ga_{0}+\Ga)\ro_{es}-\left[\, (\Ga
+i\Omega)+2(\Ga_{0}+2i\om)\,\right]\ro_{sg}\label{eq10}
\end{eqnarray}
\end{subequations}
and for the remaining matrix elements one can use hermiticity of
$\ro$. Notice that in (\ref{eq}), equations (\ref{eq1}) --
(\ref{eq4}), (\ref{eq5}) -- (\ref{eq6}) and (\ref{eq7}) --
(\ref{eq10}) are decoupled and can be solved independently.
Observe also that four -- level system prepared in symmetric state
$\ket{s}$ decays with enhanced  rate $\Ga_{0}+\Ga$, whereas
antisymmetric initial state $\ket{a}$ leads to reduced  rate
$\Ga_{0}-\Ga$. When two atoms are confined in a region smaller
than the resonant wave length, we can put $\Ga=\Ga_{0}$ (see e.g.
\cite{FT}) so antisymmetric state $\ket{a}$ is completely
decoupled from the environment and is decoherence -- free state of
the two -- atomic system.
\par
One can also check that equations (\ref{eq}) describe two types of
time evolution of the system, depending on the relations between
$\Ga$ and $\Ga_{0}$. When $\Ga<\Ga_{0}$, there is a unique
asymptotic state of the system, which is maximally mixed state
$\frac{\DS \I_{4}}{\DS 4}$. In that case the relaxation process
brings all initial states of two atoms into the state with maximal
entropy. In the present note, we analyse the case when
$\Ga=\Ga_{0}$ and show that asymptotic stationary states depend on
initial conditions and can be parametrized by matrix elements
$\ro_{aa}$ of the initial states.
\section{Small distance separation and generation of Werner states}
When $\Ga=\Ga_{0}$, equations (\ref{eq}) simplify considerably and
one can check that solutions of (\ref{eq5}) -- (\ref{eq10})
asymptotically vanish. Nontrivial contribution to the asymptotic
stationary states comes from the matrix elements $\ro_{aa},\,
\ro_{ss},\, \ro_{ee}$ and $\ro_{gg}$ satisfying equations
(\ref{eq1}) -- (\ref{eq4}), which in that case can be written as
follows
\begin{eqnarray}
\frac{d\ro_{aa}}{dt}&&=0\label{as1}\\
\frac{d\ro_{ss}}{dt}&&=2\Ga_{0}\,(\ro_{ee}+\ro_{gg}-2\ro_{ss})\label{as2}\\
\frac{d\ro_{ee}}{dt}&&=2\Ga_{0}\,(\ro_{ss}-\ro_{ee})\label{as3}\\
\frac{d\ro_{gg}}{dt}&&=2\Ga_{0}\,(\ro_{ss}-\ro_{gg})\label{as4}
\end{eqnarray}
The system of equations (\ref{as1}) -- (\ref{as4}) can be solved and
we obtain
\begin{eqnarray}
\ro_{aa}(t)&&=\ro_{aa}(0)\\
\ro_{ss}(t)&&=\frac{1}{3}\, (1-\ro_{aa}(0))+\frac{1}{3}\,
e^{-6\Ga_{0}t}\, (\ro_{aa}(0)+3\ro_{ss}(0)-1)\\
\ro_{ee}(t)&&=\frac{1}{3}\,(1-\ro_{aa}(0))+\frac{1}{6}\,
e^{-6\Ga_{0}t}\,(1-\ro_{aa}(0)-3\ro_{ss}(0))+\frac{1}{2}\,e^{-2\Ga_{0}t}\,
(\ro_{aa}(0)+\ro_{ss}(0)+2\ro_{ee}(0)-1)\\
\ro_{gg}(t)&&=\frac{1}{3}\,(1-\ro_{aa}(0))+\frac{1}{6}\,
e^{-6\Ga_{0}t}\,(1-\ro_{aa}(0)-3\ro_{ss}(0))+\frac{1}{2}\,e^{-2\Ga_{0}t}\,
(\ro_{aa}(0)+\ro_{ss}(0)+2\ro_{gg}(0)-1)
\end{eqnarray}
Thus
$$
\lim\limits_{t\to\infty} \ro_{ss}(t)=\frac{1}{3}\,(1-\ro_{aa}(0))
$$
and similarly for $\ro_{ee}(t)$ and $\ro_{gg}(t)$. So the stationary
asymptotic states $\ro_{\infty}$ are parametrized by
$\ro_{aa}(0)=\bra{a}\ro\ket{a}$, where $\bra{a}\ro\ket{a}=F$ is the
\textit{fidelity} of the initial state $\ro$ with respect to the
state $\ket{a}$ (or the overlap of $\ro$ with singlet state
$\ket{a}$). In the canonical basis (\ref{canbasis}), $\ro_{\infty}$
has the form
\begin{equation}
\ro_{\infty}=\begin{pmatrix} \frac{\DS 1-F}{\DS 3}&0&0&0\\[2mm]
0&\frac{\DS 1+2F}{\DS 6}&\frac{\DS 1-4F}{\DS 6}&0\\[2mm]
0&\frac{\DS 1-4F}{\DS 6}&\frac{\DS 1+2F}{\DS 6}&0\\[2mm]
0&0&0&\frac{\DS 1-F}{\DS 3}
\end{pmatrix}\label{asstate}
\end{equation}
Notice that for some values of parameter $F$, the state
$\ro_{\infty}$ is entangled. If we compute its concurrence given by
the well known formula \cite{HW, Woo}
$$
C(\ro)=\max\, (0,\, 2\lambda_{\mr{max}}(\widehat{\ro})-\tr
\widehat{\ro}\,)
$$
where $\lambda_{\mr{max}}(\widehat{\ro})$ is the maximal eigenvalue
of $\widehat{\ro}$ and
$$
\widehat{\ro}=\sqrt{\sqrt{\ro}\,\widetilde{\ro}\,\sqrt{\ro}},\quad
\widetilde{\ro}=(\si{2}\otimes\si{2})\overline{\ro}(\si{2}\otimes\si{2})
$$
with $\overline{\ro}$ denoting complex conjugation of the matrix
$\ro$, then we obtain
$$
C(\ro_{\infty})=\Wasy{0}{F\leq 1/4}{2F-1}{F>1/4}
$$
So we see that for \textit{any initial state} $\ro$ with fidelity
$F$, there exist asymptotic state $\ro_{\infty}$ such that:\\[2mm]
{\textbf{1.}} if $F\in [0,1/4),\; \ro_{\infty}$ is separable and
can be written as
\begin{equation}
\ro_{\infty}=\frac{1}{4}\,\begin{pmatrix} 1+\frac{\DS p}{\DS
3}&0&0&0\\[2mm]
0&1-\frac{\DS p}{\DS 3}&\frac{\DS 2p}{\DS 3}&0\\[2mm]
0&\frac{\DS 2p}{\DS 3}&1-\frac{\DS p}{\DS 3}&0\\[2mm]
0&0&0&1+\frac{\DS p}{\DS 3}
\end{pmatrix},\quad p=1-4F\label{assepar}
\end{equation}
\vskip 2mm\noindent
{\textbf{2.}} if $F=\frac{1}{4},\;
\ro_{\infty}=\frac{\DS\I_{4}}{\DS 4}$\\[2mm]
{\textbf{3.}} if $F\in (1/4,1],\; \ro_{\infty}$ is equal to the
Werner state
\begin{equation}
W_{a}=(1-p)\,\frac{\DS \I_{4}}{\DS
4}+p\,\ket{a}\bra{a}\label{wernerstate}
\end{equation}
\hspace*{3mm} with $p=\frac{\DS 4F-1}{\DS 3}$. It is separable for
$F\in (1/4,1/2]$ and entangled for $F>1/2$ with concurrence
$C(\ro_{\infty})=2F-1$.\\[4mm]
This result shows that all initially entangled states with
fidelity greater then $1/2$ preserve some amount of their
entanglement during the interaction with maximally noisy
environment and evolve into Werner states with the same fidelity.
The class of Werner states has interesting properties: they
interpolate between maximally entangled and maximally mixed
states, for that class it was shown that some mixed entangled
states can satisfy Bell inequalities \cite{W}, they have maximal
possible entanglement with respect to the non -- local unitary
transformations and local and non -- local general operations
\cite{HI}, Werner states can be also applied in entanglement
teleportation via mixed states \cite{LK}.
\par
Let us stress that the asymptotic behavior of the initial state
depends only on its overlap with  the singlet state $\ket{a}$ and
not on its entanglement. There are many states with the same
entanglement for which fidelity varies from $0$ to maximal value.
This is for example the case of maximally entangled pure states. So
initial states with the same entanglement can behave very
differently with respect to the noise. The states which are more
"similar" to the singlet state are more stable. Thus to protect the
initial entanglement we may perform appropriate local operations on
initial states to maximize its fidelity. All these aspects of noisy
dynamics will be discussed on explicit examples in the next section.
\section{Some examples}
\subsection{Pure separable initial states}
Let $\ro=\ket{\Psi\otimes \Phi}\bra{\Psi\otimes\Phi}$ for $\Psi,\,
\Phi\in \C^{2}$. Since for this state
\begin{equation}
F=\frac{1}{2}(1-|\ip{\Psi}{\Phi}|^{2})
\end{equation}
 $0\leq F\leq \frac{\DS 1}{\DS 2}$ and  all asymptotic states are
separable. Depending on the value of $|\ip{\Psi}{\Phi}|^{2}$ we
have the following
possibilities:\\[2mm]
\textbf{1.} if $\frac{\DS 1}{\DS 2}< |\ip{\Psi}{\Phi}|^{2}\leq 1$,
then the asymptotic state $\ro_{\infty}$ is equal to the state
(\ref{assepar}) with
$p=2|\ip{\Psi}{\Phi}|^{2}-1$,\\[2mm]
\textbf{2.} if $|\ip{\Psi}{\Phi}|^{2}=\frac{\DS 1}{\DS 2}$, then
$\ro_{\infty}=\frac{\DS\I_{4}}{\DS 4}$\\[2mm]
\textbf{3.} if $0\leq |\ip{\Psi}{\Phi}|^{2}<\frac{\DS 1}{\DS 2}$,
then $\ro_{\infty}$ is equal to separable Werner state $W_{a}$ given
by (\ref{wernerstate}) with $p=\frac{\DS
1-2|\ip{\Psi}{\Phi}|^{2}}{\DS 3}$. Notice that in contrast to the
zero temperature case, where purely incoherent dissipative process
can lead to the creation of entanglement \cite{J,FT1,FT2,JJ} in the
present model initial separable states remain separable.
\subsection{Pure maximally entangled initial states}
Consider now the class of maximally entangled states \cite{BJO}
\begin{equation}
P(a,\theta_{1},\theta_{2})=\frac{1}{2}\,\begin{pmatrix}\kw&a\sqrt{\aaa}e^{-i\theta_{1}}
&a\sqrt{\aaa}e^{-i\theta_{2}} &-\kw
e^{-i(\theta_{1}+\theta_{2})}\\[2mm]
a\sqrt{\aaa} e^{i\theta_{1}}&\aaa&(\aaa)
e^{i(\theta_{1}-\theta_{2})}&-a\sqrt{\aaa} e^{-i\theta_{2}}\\[2mm]
a\sqrt{\aaa} e^{i\theta_{2}}&(\aaa)
e^{-i(\theta_{1}-\theta_{2})}&\aaa&
-a\sqrt{\aaa} e^{-i\theta_{1}}\\[2mm]
-\kw e^{i(\theta_{1}+\theta_{2})}&-\sqrt{\aaa}
e^{i\theta_{2}}&-a\sqrt{\aaa} e^{i\theta_{1}}&\kw
\end{pmatrix}\label{maksent}
\end{equation}
where $a\in [0,1],\; \theta_{1},\theta_{2}\in [0,2\pi]$. All states
from the class (\ref{maksent}) have  concurrence equal to $1$. On
the other hand
\begin{equation}
F=\frac{1}{2}\, (1-\kw)(1-\cos (\theta_{1}-\theta_{2}))\label{fid}
\end{equation}
One can check that fidelity $F$ can take all values from $0$ to
$1$ depending on parameters $a$ and
$\theta=\theta_{1}-\theta_{2}$. In particular $F>\frac{1}{2}$
inside the set $E$ on the $(a,\theta)$ - plane, given by
\begin{equation}
E=\{ (a,\theta)\,:\, 0\leq a\leq \frac{1}{\sqrt{2}},\; \arccos
\frac{a^{2}}{a^{2}-1}<\theta<2\pi -\arccos\frac{a^{2}}{a^{2}-1}
\}\label{set}
\end{equation}
Outside this set, $F<\frac{1}{2}$ . On the curve
\begin{equation}
\theta=\arccos\frac{2a^{2}-1}{2(a^{2}-1)},\; a\in
[0,\sqrt{3}/2]\label{curve}
\end{equation}
\begin{figure}
\centering
\includegraphics[width=80mm]{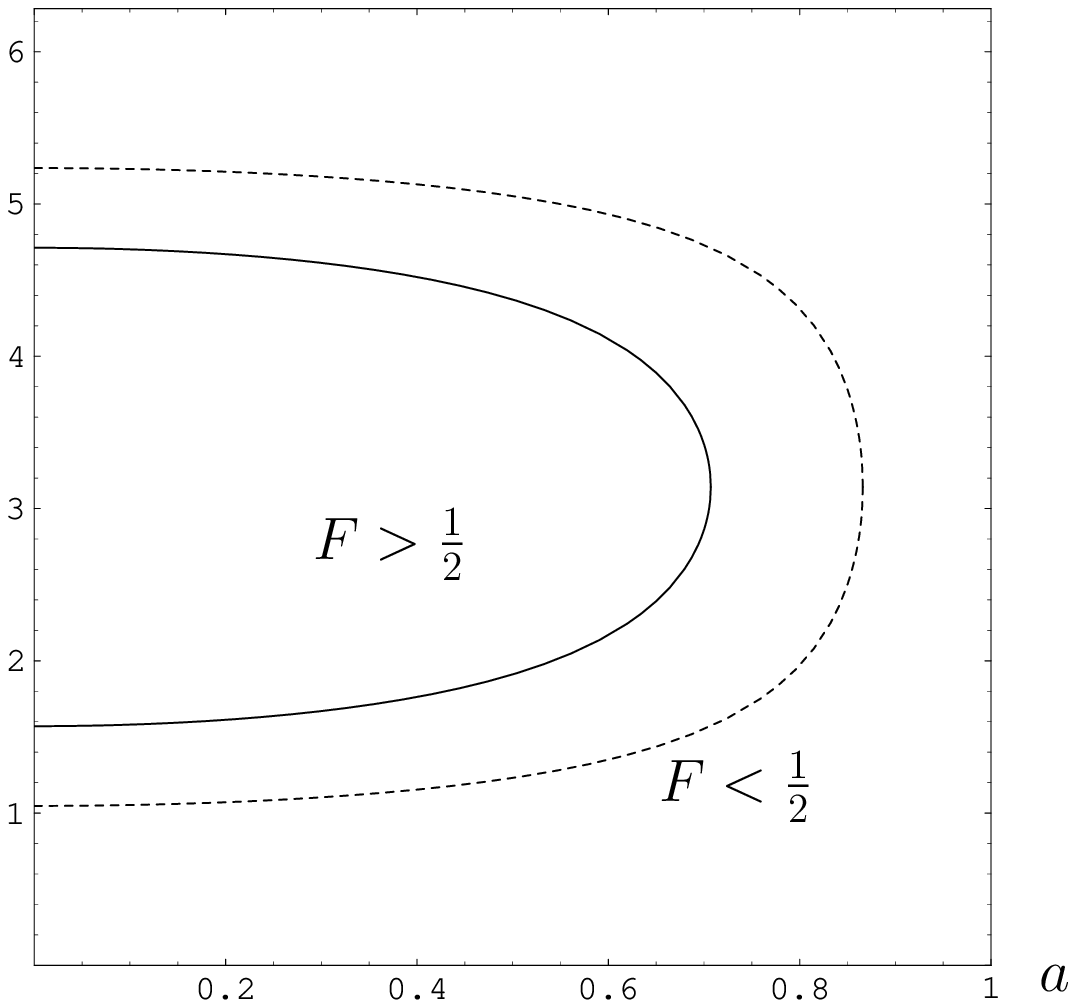}
\caption{Fidelity of maximally entangled states as function of $a$
and $\theta_{1}-\theta_{2}$. On the dotted curve $F=\frac{1}{4}$}
\end{figure}
the fidelity is equal to $\frac{1}{4}$ (see FIG.1). Thus all
initial states $P(a,\theta_{1},\theta_{2})$ with
$(a,\theta_{1}-\theta_{2})\in E$ evolve into  entangled Werner
states, whereas states with $(a,\theta_{1}-\theta_{2})$ outside
$E$ become separable. When these parameters lie on the curve
(\ref{curve}), the dynamics brings corresponding initial maximally
entangled states into maximally mixed state $\frac{\DS \I_{4}}{\DS
4}$. For all initial states $P(a,\theta_{1},\theta_{2})$ with
$(a,\theta_{1}-\theta_{2})\in E$, asymptotic concurrence is
smaller then $1$, except antisymmetric state $\ket{a}$ which is
stable.

\subsection{Some mixed initial states}
\subsubsection{Mixed separable states}
If $\ro$ is mixed separable state, then it can be written as
\begin{equation}
\ro=\sum\limits_{k}s_{k}P_{k},\quad s_{k}\geq 0,\,
\sum\limits_{k}s_{k}=1\label{mixsep}
\end{equation}
where $P_{k}$ are pure separable states. Since for any pure
separable state, fidelity is not greater then $1/2$, by
(\ref{mixsep}) the same is true for all mixed separable states. So
they evolve into separable asymptotic states.
\subsubsection{Werner states}
Let
$$
\ket{\pm}=\frac{1}{\sqrt{2}}(f_{1}\pm f_{4})
$$
Besides $W_{a}$ define also the states
\begin{equation}
W_{s}=(1-p)\frac{\I_{4}}{4}+p\,\ket{s}\bra{s},\quad
W_{\pm}=(1-p)\frac{\I_{4}}{4}+p\,\ket{\pm}\bra{\pm}\label{werner3}
\end{equation}
One can check that for all states (\ref{werner3})
$$
F=\frac{1-p}{4}
$$
so $0\leq F<\frac{\DS 1}{\DS 4}$ and they evolve to asymptotic
separable state (\ref{assepar}). On the other hand, the states
(\ref{werner3}) are locally equivalent to $W_{a}$. If we define
$$
U_{s}=\si{3}\otimes \I_{2},\quad U_{+}=\I_{2}\otimes i\si{2},\quad
U_{-}=\I_{2}\otimes \si{1}
$$
then
$$
W_{s}=U_{s}W_{a}U_{s}^{\ast},\quad
W_{+}=U_{+}W_{a}U_{+}^{\ast},\quad W_{-}=U_{-}W_{a}U_{-}^{\ast}
$$
\subsubsection{Bell-diagonal states}
Let $\ro_{B}$ be the convex combination of pure states $\ket{+},\,
\ket{-},\, \ket{s}$ and $\ket{a}$
\begin{equation}
\ro_{B}=p_{1}\ket{+}\bra{+}+p_{2}\ket{-}\bra{-}+p_{3}\ket{s}\bra{s}
+p_{4}\ket{a}\bra{a}\label{belldiag}
\end{equation}
It is known that if all $p_{i}\in [0,1/2],\, \ro_{B}$ is
separable, while for $p_{1}>1/2,\, \ro_{B}$ is entangled with
concurrence equal to $2p_{1}-1$ (similarly for $p_{2},\, p_{3},\,
p_{4}$) \cite{HH}. On the other hand, for states (\ref{belldiag})
$$
F=p_{4}
$$
so all states (separable or entangled) with $p_{4}<1/2$ become
separable asymptotically. If $p_{4}>1/2$ then noisy dynamics
produces asymptotic Werner state $W_{a}$ with concurrence equal to
$2F-1=2p_{4}-1$. In this case asymptotic entanglement is exactly
equal to initial entanglement, so the amount of entanglement is
preserved. In the next section, we discuss this interesting
phenomenon for some class of initial states.
\section{Preservation of entanglement for some non-maximally
entangled initial states}
Let us discuss now how entanglement of
the asymptotic state can depend on initial entanglement. This
problem has a simple solution in the case when initial
entanglement is a function of fidelity $F$. Since $F$ is constant
during the evolution, the final entanglement is exactly equal to
its initial value. So the process of collective damping can
preserve entanglement of some initial states. Simple examples of
such states are described below. Let
\begin{equation}
\ro=\begin{pmatrix}
0&0&0&0\\
0&\ro_{22}&\ro_{23}&0\\
0&\ro_{23}&\ro_{33}&0\\
0&0&0&0
\end{pmatrix}\label{inistate}
\end{equation}
Notice that in (\ref{inistate}) all matrix elements are real and
$F=\frac{1}{2}\,(1-2\ro_{23})$. Moreover
\begin{equation}
C(\ro)=2|\ro_{23}|=|1-2F|
\end{equation}
From our previous results it follows that:\\[2mm]
\textbf{a.} if $\ro_{23}\geq 0$ i.e. $F\leq \frac{1}{2}$, then
$C(\ro)=2\ro_{23}\geq 0$, but $C(\ro_{\infty})=0$,\\[2mm]
\textbf{b.} if $\ro_{23}<0$ i.e. $F>\frac{1}{2}$, then
$C(\ro)=-2\ro_{23}=2F-1$ and $C(\ro_{\infty})=2F-1$, so
\begin{equation}
C(\ro_{\infty})=C(\ro)\label{entinias}
\end{equation}
Notice that the unitary operator $\si{3}\otimes \I_{2}$ transforms
the states (\ref{inistate}) with $\ro_{23}>0$ to the states with
$\ro_{23}<0$. So performing local operations on the initial state,
we can protect its entanglement.
\par
Consider also explicit examples of states for which the relation
(\ref{entinias}) holds. Let us take the class of pure states
\begin{equation}
\Psi=\cos\phi\, \ket{0}\otimes \ket{1}+\sin\phi\,
\ket{1}\otimes\ket{0},\quad \phi\in [0,\pi]\label{pure}
\end{equation}
One can check that for the class (\ref{pure})
$$
F=\frac{1}{2}\,(1-\sin 2\phi)
$$
and
$$
C(\ro_{\infty})=\max\, (0,-\sin 2\phi )
$$
We see that all pure entangled states (\ref{pure}) with $\phi\in
[\pi/2,\pi]$, evolve into asymptotic Werner states, which have the
same entanglement as initial states.
\par
It is instructive to discuss  evolution of initial states
(\ref{inistate}) also for finite times. Here the dipol -- dipol
interaction between atoms plays a significant role. For small
separation of atoms the ratio $\Omega/\Ga_{0}$ can be large
\cite{FT, KW} and this coupling can induce transient increase (or
decrease) of entanglement. But noise reduces the amplitude of
those oscillations of concurrence and they vanish asymptotically.
To see some details, we solve the equations (\ref{as1})--
(\ref{as4}) and (\ref{eq6}) for such initial states and obtain
\begin{equation}
\ro(t)=\begin{pmatrix} \ro_{11}(t)&0&0&0\\[2mm]
0&\ro_{22}(t)&\ro_{23}(t)&0\\[2mm]
0&\ro_{32}(t)&\ro_{33}(t)&0\\[2mm]
0&0&0&\ro_{44}(t)\end{pmatrix}\label{rot}
\end{equation}
where
\begin{subequations}
\begin{eqnarray}
\ro_{11}(t)&&=\frac{1}{6}(1+2\ro_{23})-\frac{1}{6}e^{-6\Ga_{0}t}(1+2\ro_{23})\\
\ro_{22}(t)&&=\frac{1}{3}(1-\ro_{23})+\frac{1}{6}e^{-6\Ga_{0}t}(1+2\ro_{23})
+\frac{1}{2}e^{-2\Ga_{0}t}\cos 2\Omega t (\ro_{22}-\ro_{33})\\
\ro_{33}(t)&&=\frac{1}{3}(1-\ro_{23})+\frac{1}{6}e^{-6\Ga_{0}t}(1+2\ro_{23})
-\frac{1}{2}e^{-2\Ga_{0}t}\cos 2\Omega t (\ro_{22}-\ro_{33})\\
\ro_{44}(t)&&=\frac{1}{6}(1+2\ro_{23})-\frac{1}{6}e^{-6\Ga_{0}t}
(1+2\ro_{23})\\
\ro_{23}(t)&&=\frac{1}{6}(4\ro_{23}-1)+\frac{1}{6}e^{-6\Ga_{0}t}(1+2\ro_{23})
-\frac{i}{2}e^{-2\Ga_{0}t}\sin 2\Omega t (\ro_{22}-\ro_{33})
\end{eqnarray}
\end{subequations}
One can check that for states (\ref{rot}), concurrence is given by
\begin{equation}
C(\ro(t))=\max \left(0,\,
2\left(\,|\ro_{23}(t)|-\sqrt{\ro_{11}(t)\ro_{44}(t)}\,\right)\right)=
\max \left(0,\,
2\left(\,|\ro_{23}(t)|-\ro_{11}(t)\,\right)\right)\label{crot}
\end{equation}
so
\begin{equation}
C(\ro(t))=\max\, \left(0,\: 2\sqrt{A(t)^{2}+B(t)^{2}\sin^{2}
2\Omega t}-2\ro_{11}(t)\,\right)\label{conct1}
\end{equation}
with
\begin{equation}
A(t)=\frac{1}{6}\,
\left(4\ro_{23}-1+e^{-6\Ga_{0}t}\,(1+2\ro_{23})\right),\quad
B(t)=\frac{1}{2}e^{-2\Ga_{0}t}\, (\ro_{22}-\ro_{33})\label{conct2}
\end{equation}
This shows that the concurrence as a function of time oscillates
with period depending on the strength of dipol -- dipol
interaction but when $\ro_{23}<0$, it tends to its initial value
(FIG. 2). When the dipol -- dipol interaction is absent or when
the initial state is such that $\ro_{22}=\ro_{33}=\frac{1}{2}$ and
$\ro_{23}<0$, then from (\ref{conct1}) and (\ref{conct2}) it
follows that
\begin{equation}
|\ro_{23}(t)|-\ro_{11}(t)=-\ro_{23}(t)-\ro_{11}(t)\label{diff}
\end{equation}
Since time-dependent terms in (\ref{diff}) cancel, it is equal to
$-2\ro_{23}$, and
$$
C(\ro(t))=C(\ro)
$$
for all $t$. When $\ro_{23}>0$, then
\begin{equation}
|\ro_{23}(t)|-\ro_{11}(t)=\ro_{23}(t)-\ro_{11}(t)\label{diff1}
\end{equation}
and (\ref{diff1})depends on $t$ in such a way that $C(\ro(t))$
monotonically goes to zero.
\begin{figure}
\centering
\includegraphics[width=80mm]{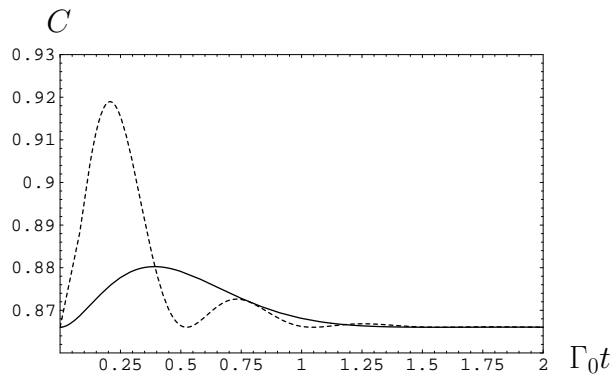}
\caption{Concurrence as function of time for initial state (\ref{pure}) with
$\phi=2\pi/3$ for $\Omega/\Gamma_{0}=1$ (solid line) and
 $\Omega/\Gamma_{0}=3$ (dotted line)}
\end{figure}
The effect of preservation of entanglement for states (\ref{rot})
should be contrasted with the monotonic decreasing of purity of
the state $\ro$ defined by $\tr \ro^{2}$, during the time
evolution (\ref{mastereq}). As shown in {\rm{\cite{Alicki}}}, it
is equivalent to the condition
$$
L_{\mr{N}}(\I_{4})=0
$$
which can be simply checked to be true in our model.
\section{Conclusions}
During the evolution of open quantum systems interacting with
environment the process of decoherence usually results in
degradation of entanglement. To preserve as much entanglement as
possible one has to control the effects of noise. In this context,
we have studied the model of compound system of two atoms
influenced by so called maximal noise, which can be treated as the
limiting case of thermal noise, when temperature goes to infinity.
As we have shown, even in that  case there are entangled states
which are decoherence -- free. Explicit examples are given by
singlet state $\ket{a}$ and Werner state $W_{a}$. On the other
hand, there are evolving states with a very interesting property:
its asymptotic entanglement is exactly equal to the initial one,
or even is stable during the time evolution. We have also shown
that performing some local operations on initial states can help
with protecting entanglement.
\begin{acknowledgments}
The authors would like to acknowledge financial support by Polish
Ministry of Scientific Research and Information Technology under the
grant PBZ-Min-008/PO3/2003.
\end{acknowledgments}

\end{document}